\documentclass[pra,twocolumn,showpacs]{revtex4}
\usepackage{graphicx}
\begin{document}

\bibliographystyle{unsrt}

%%%%%%%%%%%%%%%%%%%%%%%%%%%%%%%%%%%%%%%%%
\title{Experimental Demonstration of a Quantum Circuit using Linear Optics Gates}
\author{T.B. Pittman, B.C Jacobs, and J.D. Franson}
\affiliation{Johns Hopkins University,
Applied Physics Laboratory, Laurel, MD 20723}
%%%%%%%%%%%%%%%%%%%%%%%%%%%%%%%%%%%%%%%%%

\date{\today}

%%%%%%%%%%%%%%%%%%%%%%%%%%%%%%%%%%%%%%%%%
\begin{abstract}
One of the main advantages of an optical approach to quantum computing is the fact that optical fibers can be used to connect the logic and memory devices to form useful circuits, in analogy with the wires of a conventional computer.  Here we describe an experimental demonstration of a simple quantum circuit of that kind in which two probabilistic exclusive-OR (XOR) logic gates were combined to calculate the parity of three input qubits.
\end{abstract}
%%%%%%%%%%%%%%%%%%%%%%%%%%%%%%%%%%%%%%%%%


\maketitle

One of the most attractive features of an optical approach to quantum information processing \cite{knill01a,pittman02a,pittman03b,obrien03,sanaka04,gasparoni04,zhao04} is the ability to transport single-photon qubits between distant locations \cite{gisin02}.  This would allow individual quantum logic gates and memory devices to be connected using optical fibers or waveguides to form more complex circuits, in analogy with the use of wires in electronic circuits \cite{nielsenchuangbook}.  Here we report an experimental demonstration of the use of optical fibers to form a small-scale quantum circuit in which two exclusive-OR (XOR) quantum logic gates were combined to calculate the parity of three input qubits.

\begin{figure}[b]
\includegraphics[angle=-90,width=3in]{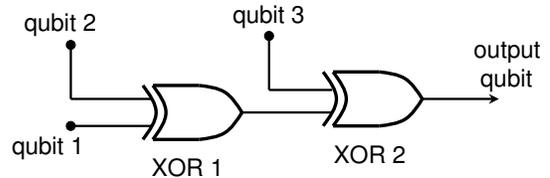}
\vspace*{-1in}
\caption{Simple logic circuit for photonic qubits.  The circuit calculates the parity of three input qubits using two probabilistic XOR logic gates that were implemented using LOQC techniques.}
\label{fig:xorcircuit}
\end{figure}

    The circuit to be implemented consists of two XOR gates in series, as illustrated in Figure \ref{fig:xorcircuit}.  Two photonic qubits, $q_{1}$ and $q_{2}$, form the input to the first XOR gate.  The output of that gate forms the input to the second XOR gate, along with a third qubit $q_{3}$ . It can be easily shown that the output of this circuit corresponds to the parity of the three input qubits. In our approach, the logical values of the qubits are represented by the linear polarization states of single photons, with a horizontal polarization corresponding to a logical value of 0 and a vertical polarization corresponding to a value of 1 \cite{koashi01}. The use of multiple (ie, three) single-photon qubits in a quantum circuit is primarily what distinguishes the present work from earlier experiments involving a single photon (see, for example, \cite{cerf98,takeuchi00}).

\begin{figure}[t]
\hspace*{-.1in}
\includegraphics[angle=-90,width=3.5in]{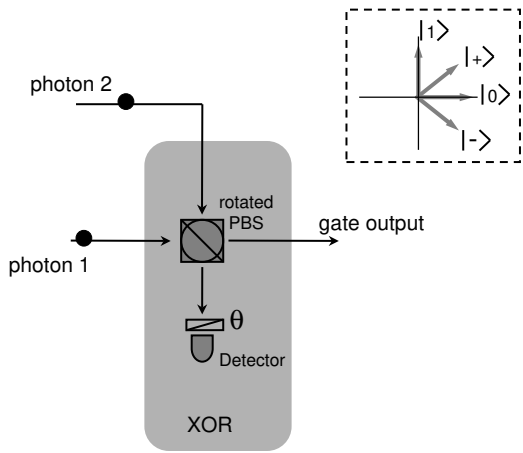}
\caption{Implementation of a probabilistic XOR gate for single-photon qubits using a single polarizing beam splitter (PBS), a polarization analyzer $\theta$, and a single-photon detector \protect\cite{pittman01a}. The dashed-box inset shows the polarization conventions used throughout the text. The PBS is oriented to transmit polarization states corresponding to $|+\rangle$ and reflect polarization states corresponding to $|-\rangle$. }
\label{fig:dcnot}
\end{figure}

    The XOR logic gates were implemented using linear optics quantum computing (LOQC) techniques, as first suggested by Knill, Laflamme, and Milburn \cite{knill01a}.  Although logic operations are inherently nonlinear, they showed that probabilistic logic operations could be implemented using linear optical elements, additional photons (ancilla), and post-selection and feed-forward control based on the results of measurements.  Our implementation of a single probabilistic XOR logic gate using polarization encoding and linear optical elements is illustrated in Fig. \ref{fig:dcnot}.  The device consists of a polarizing beam splitter oriented at an angle of $45^{o}$ to the vertical, a polarization analyzer oriented at an angle $\theta$, and a single-photon detector.  The analyzer is oriented at an angle $\theta=0^{o}$  during normal operation and the output of the device is only accepted if the detector registers one and only one photon, which occurs with a probability of $P_{s}$  to be described below.  The success rate can be doubled by using two detectors and a second polarizing beam splitter combined with feed-forward control \cite{pittman02b}, but in this experiment we used only a single detector in each XOR gate.

The most general input state to a single XOR gate has the form:

\begin{equation}
|\psi_{0}\rangle = \alpha|0,0\rangle +\beta|0,1\rangle +\gamma|1,0\rangle +\delta|1,1\rangle
\label{eq:inputstate}
\end{equation}

\noindent We previously showed that, when a single photon is detected, this device will implement a non-unitary transformation given by \cite{pittman01a,pittman02a}:

\begin{equation}
|\psi_{0}\rangle \rightarrow \alpha|0\rangle +\beta|1\rangle +\gamma|1\rangle +\delta|0\rangle
\label{eq:outputstate}
\end{equation}

%%%%%%%%%%%%%%%%%%%%%%%%%%%%%%%%%%
%%%%%%%%%%%%%%%%%%%%%%%%%%%%%%%%%%%

%%%%%%%%%%%%%%%%%%%%%%%%%%%%%%%%%%
%%%%%%%%%%%%%%%%%%%%%%%%%%%%%%%%%%%

It can be seen that the values of the two input qubits have been erased (in the sense of a quantum eraser) and that the output of the device corresponds to that of an XOR gate, as desired.

The probability $P_{s}$ that the operation will succeed is given by:

\begin{equation}
P_{s}= \frac{1}{4}\left( \left|\alpha+\delta \right|^{2}+ \left| \beta+ \gamma \right|^{2} \right)
\label{eq:probabilityofsuccess}
\end{equation}

\noindent Unlike most linear optics gates, here the value of $P_{s}$ depends on the input state and it vanishes for certain superposition states.  When the gate succeeds, however, it does produce the correct superposition of output states.  Although this feature could limit the usefulness of XOR gates of this kind, in most applications of interest the input state does not consist of a fixed superposition state and corresponds instead to an entangled or mixed state.  For example, these devices can be used as part of a full CNOT gate that succeeds with a probability of $\frac{1}{4}$ for an arbitrary input state \cite{pittman01a}. It should also be noted that XOR gates are not reversible. Here we are primarily interested in the possiblity of constructing simple circuits rather than the properties of the gates themselves.

\begin{figure}[b]
\vspace*{-.3in}
\hspace*{-.1in}
\includegraphics[angle=-90,width=3.5in]{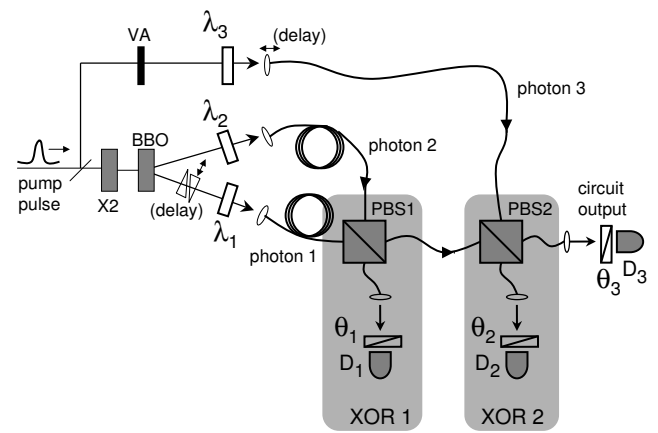}
\caption{Experimental apparatus used to demonstrate the fiber-based circuit of two probabilistic XOR gates. The two XOR gates are represented by the two shaded regions, and the three input photons were derived from a parametric down-conversion source and a weak coherent state pulse.  Additional details and symbols are described in the text.}
\label{fig:experiment}
\end{figure}

The circuit shown in Figure \ref{fig:xorcircuit} requires three single-photon qubits and two XOR gates as described above.  In order to demonstrate this circuit, we used a three-photon source in which two of the necessary single-photon qubits were produced by parametric down-conversion while the third qubit was obtained from the attenuated pump laser pulse \cite{rarity97,pittman03a}. This required multi-photon experimental techniques similar to those used, for example, in the original quantum teleportation experiments \cite{zukowski95,bouwmeester97}. An overview of the experimental apparatus is shown in Figure \ref{fig:experiment}.   

The experiment was driven by a mode-locked Ti-Sapphire laser providing short pump pulses ($\sim$150fs, 780nm) at a repetition rate of 76MHz. These pulses were frequency doubled to the UV (crystal X2, 390nm) and used to pump a down-conversions source (BBO, Type I cut, 0.7mm thick) that provided photon 1 and photon 2.  Photon 3 was obtained by picking off a small fraction of the original 780nm pulse and reducing it to the single-photon level with a variable attenuator (VA).  All three photons were coupled into single-mode fibers which served as the ``wires'' of the circuit.  

%%%%%%%%%%%%%%%%%%%%%%%%%%%%%%%%%%
%%%%%%%%%%%%%%%%%%%%%%%%%%%%%%%%%%%

The two shaded regions of Figure \ref{fig:experiment} highlight the two XOR gates that comprise the circuit.  Each XOR gate consisted of a fiber-connected bulk polarizing beam splitter (PBS1, PBS2), a  polarization analyzer ($\theta_{1}$,$\theta_{2}$), and a single-photon detector ($D_{1}$, $D_{2}$). The reflection/transmission axes of these PBS's defined the definitions of the polarizations states $|+\rangle$ and $|-\rangle$, and the analyzers $\theta_{1}$ and $\theta_{2}$ were fixed at an orientation corresponding to $|0\rangle$. Half-wave plates $\lambda_{i}$, (i=1,2,3), in the input beams could be used to produce superposition states of the input qubits.  
For a given choice of the three input qubits, a third rotatable polarization analyzer and detector ($\theta_{3}$ and $D_{3}$) could be used to verify the expected logical operation of the entire circuit by measuring the polarization of the output photon. 

In practice, data accumulation involved measuring a three-fold counting rate consisting of the sequential firing of the gate detectors $D_{1}$ and $D_{2}$, followed by the output detector $D_{3}$. A narrow electronic coincidence window was defined to ensure that all three detected photons corresponded to the same original pumping pulse.  This type of ``coincidence basis'' operation was required to overcome the problems associated with the random nature of the photon sources, the various optical losses in the system, and the inability of the detectors to resolve photon-number.  In this arrangement,  the largest source of noise was due to events in which one (or two) down converted photons triggered $D_{1}$ while events at the other two detectors were due to the small probability that the weak coherent state pulse actually contained two photons. The magnitude of the weak coherent state was therefore adjusted to keep the ratio of error events to valid three-photon events on the order of $10^{-2}$ \cite{pittman03a}.

\begin{figure}
\hspace*{-.3in}
\includegraphics[angle=-90,width=3.9in]{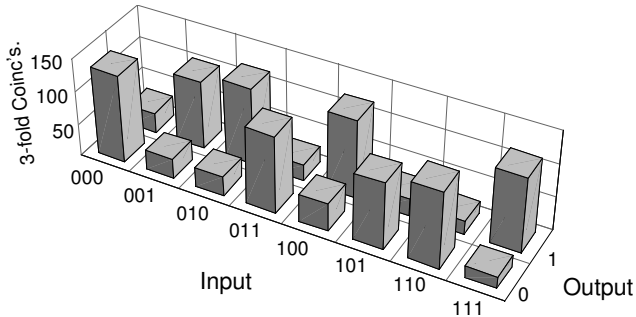}
\vspace*{-1in}
\caption{Experimental results obtained for all possible combinations of the basis states $|0\rangle$ and $|1\rangle$, which corresponds to the classical truth table for the circuit. For each of the eight possible combinations of input basis states, the data show the number of three-fold coincidence counts per 1200 seconds obtained with $\theta_{3}$ corresponding to the output values $|0\rangle$ and $|1\rangle$. The data corresponds to an overall circuit error rate of roughly 19\%.}
\label{fig:basisstatedata}
\end{figure}

Figure \ref{fig:basisstatedata} shows the experimental results obtained when each of the input qubits was prepared in one of the basis states, $|0\rangle$ or $|1\rangle$.  It can be seen that these results correspond to the classical truth table for the circuit aside from experimental errors on the order of 19\%.  These errors were due primarily to the lack of indistinguishability between photons from down-conversion and that obtained from the pump laser pulse.  The use of narrowband spectral filters would increase the indistinguishability of the photons and would be expected to reduce the magnitude of these errors \cite{zukowski95}.

%%%%%%%%%%%%%%%%%%%%%%%%%%%%%%%%%%
%%%%%%%%%%%%%%%%%%%%%%%%%%%%%%%%%%% 

\begin{figure}[b]
\hspace*{-.2in}
\includegraphics[angle=-90,width=3.6in]{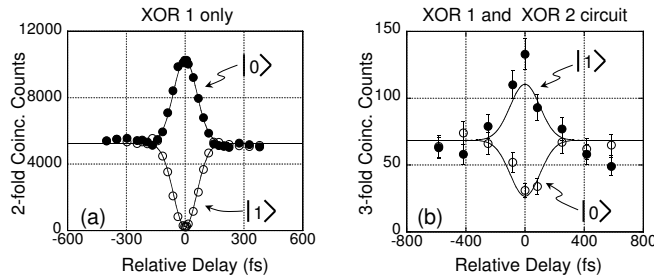}
\vspace*{-1.25in}
\caption{Experimental data demonstrating the dependence of the circuit on the spatial overlap of the photon wavepackets at a polarizing beam splitter. For this data, the input state was chosen to be $|0,0,1\rangle$. (a) shows the output obtained by temporarily interrupting the circuit after XOR1. The data shows the number of two-fold coincidence counts (per 60 seconds) as a function of the relative arrival time of photons 1 and 2 at PBS1. The solid points correspond to a logical output of $|0\rangle$ while the open circles correspond to the logical output value $|1\rangle$.  The data was fit to a Gaussian envelope function with a visibility of (95.9 $\pm$ 0.1)\%. (b) shows the output of the entire circuit. This data shows the number of three-photon coincidence counts (per 1200 seconds) as a function of the relative delay imposed on input photon 3.  Here the solid points correspond to a logical output of $|1\rangle$ while the open circles correspond to the logical output value $|0\rangle$. The data was fit to a Gaussian envelope with a visibility of (61.1 $\pm$ 6.3)\%. }
\label{fig:peakdips}
\end{figure}

\begin{figure}[t]
\vspace*{-.4in}
\hspace*{-.2in}
\includegraphics[angle=-90,width=3.5in]{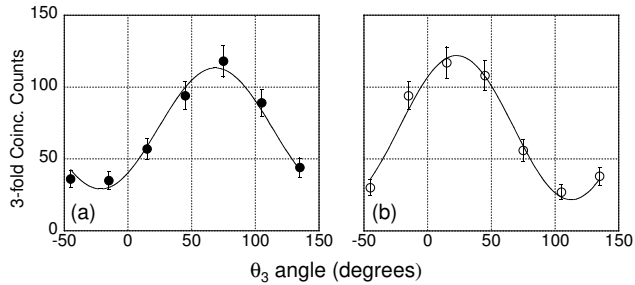}
\vspace*{-.75in}
\caption{Experimental results obtained when qubit 1 was prepared in a superposition state.  The data show the number of three-fold coincidence counts (per 1200 seconds) as a function of the output analyzer $\theta_{3}$ setting relative to $|0\rangle$.  (a) shows the results obtained when the three input qubits were in the state $|15^{o},0,1\rangle$ while (b) shows the results obtained for the input state $|15^{o},1,1\rangle$. In each case, the solid lines are sinusoidal fits to the data.  The slight shift away from the expected output polarization state ($|75^{o}\rangle$ in (a), $|15^{o}\rangle$ in (b)) was primarily due to small uncompensated birefringences in the optical fibers.  }
\label{fig:superpositiondata}
\end{figure}

    One counter-intuitive feature of this particular circuit implementation is that the basis-state examples rely most heavily on two and three-photon quantum interference effects.  Nonetheless, these basis-state examples could in principle be reproduced by an entirely classical device which, for example, destructively measured the input values and then reproduced a single output photon in the correct logical state. However, such a classical device would fail to produce the correct coherent output if one or more of the input qubits was arbitrarily chosen to be in some more general superposition of $|0\rangle$ of $|1\rangle$.  The quantum nature of the circuit could therefore be demonstrated by using both the quantum interference effects illustrated in Figure \ref{fig:peakdips}, and the circuit's response to a superpostion state input as shown in Figure \ref{fig:superpositiondata}.

 Figure \ref{fig:peakdips} shows the variation in the output of the device as a function of the difference in path lengths, which affects the spatial overlap of the photon wave packets when they arrive at the polarizing beam splitters. For this example, the input state was chosen to be $|0,0,1\rangle$.  The data shown in Fig. \ref{fig:peakdips}(a) correspond to the output of XOR 1 before it is input to XOR 2.

%%%%%%%%%%%%%%%%%%%%%%%%%%%%%%%%%%
%%%%%%%%%%%%%%%%%%%%%%%%%%%%%%%%%%%

This was accomplished by temporarily blocking the photon 3 input, removing $\theta_{2}$ and $\theta_{3}$, and applying a calibrated stress to the fiber linking $PBS1$ and $PBS2$ to cause a $45^{o}$ rotation before $PBS2$. In this way, a joint two-fold detection between $D_{1}$ and  $D_{3}$ corresponded to a logical output of $|0\rangle$ from XOR 1, while a joint detection between $D_{1}$ and $D_{2}$ corresponded to a logical output of $|1\rangle$ from XOR 1. The data shows the number of two-fold coincidences for each of these logical outputs as a function of a relative delay imposed by two translating glass wedges in the photon 1 input path.  This can be interpreted as a polarization-based manifestation of the well-known Hong-Ou-Mandel effect \cite{hong87,pan98,bouwmeester99}. When the relative delay is much larger than the coherence time of the photons, there are no two-photon interference effects and the probability of getting the correct logical output ($|0\rangle$ in this case) is roughly equal to the probability of getting the wrong logical output ( $|1\rangle$ ) in analogy with classical coin-flipping. However, when the relative delay is zero, nearly all registered two-photon events (roughly 98\% in Figure \ref{fig:peakdips}(a)) correspond to the correct logical output, while the probability of obtaining the incorrect output is almost completely suppressed by two-photon interference. In Figure \ref{fig:peakdips}(a), the imperfect suppression of these output events is equivalent to an XOR1 error rate of roughly 2\%.

The data shown in Figure \ref{fig:peakdips}(b) represent the output of the entire circuit of two XOR gates illustrated in Figure \ref{fig:experiment}. The data show the number of three-fold coincidence events as a function of the relative delay imposed by translating the input fiber of photon 3.  The two data sets shown are for the settings of $\theta_{3}$ corresponding to the logical values of $|0\rangle$ and $|1\rangle$.   Although the visibility of the three-photon peak-dip interference pattern is less than the two-photon case, the correct operation of the entire circuit is evident in an analogous fashion: at a relative delay of zero, the probability of obtaining the correct result ($|1\rangle$ in this case) is enhanced while the probability of obtaining the incorrect result ($|0\rangle$) is suppressed by three-photon interference. Once again, the reduced three-photon visibility was primarily due to the use of interference filters with a relatively wide bandpass value (FWHM $\sim$ 10nm) in front of the detectors \cite{zukowski95}. In principle, much higher visibility (and consequently lower error rates) could be obtained by using narrower filters (see, eg. \cite{pittman03a}) at the cost of much lower counting rates.

The non-classical nature of this circuit could be further illustrated by its response to a superposition state of one or more of the input qubits.  The data shown in Fig. \ref{fig:superpositiondata} correspond to the case in which photon 1 was prepared in the state $|15^{o}\rangle \equiv .97|0\rangle + .26|1\rangle$, which corresponds to a linear polarization of roughly $15^{o}$ from $|0\rangle$. For the data shown in Figure \ref{fig:superpositiondata}(a), photons 2 and 3 were respectively prepared as $|0\rangle$ and $|1\rangle$. In this case, the full XOR circuit would be expected to produce a coherent output state of $|75^{o}\rangle \equiv .26|0\rangle + .97|1\rangle$. For the data shown in Figure \ref{fig:superpositiondata}(b), photons 2 and 3 were both prepared as $|1\rangle$.  In this case the XOR circuit would be expected to produce an output state of $|15^{o}\rangle$. In both cases, the experimental data showed Malus' law-type curves consistent with these predictions. 
  
%%%%%%%%%%%%%%%%%%%%%%%%%%%%%%%%%%
%%%%%%%%%%%%%%%%%%%%%%%%%%%%%%%%%%%

In summary, we have experimentally demonstrated a quantum circuit for photonic qubits using the techniques of LOQC \cite{knill01a}.  The circuit consisted of two probabilistic XOR gates in series and it could be used to calculate the parity of three input qubits.  The operation of this circuit relied on multi-photon interference effects, and superposition states can exist in both the input and output.
The typical error rates in this proof-of-principle experiment are larger than those associated with ion trap experiments \cite{monroe95,demarco02,schmidtkaler03,leibfried03}, for example. As we mentioned before, the error rates in linear optics experiments could be reduced using an improved source of single photons, while the ability to make simple connections between logic and memory devices is unique to an optical approach. Earlier experiments in LOQC have concentrated on the demonstration of single logic gates \cite{pittman02a,pittman03b,obrien03,sanaka04,gasparoni04,zhao04}.  The results presented here provide the first explicit demonstration of the ability to connect quantum logic gates using optical fibers to construct simple quantum circuits for photonic qubits.

This work was supported by ARO, NSA, ARDA, ONR, and IR\&D funding.

%%%%%%%%%%%%%%%%%%%%%%%%%%%%%%%

%%%%%%%%%%%%%%%%%%%%%%%%%%%%%%%%

\end{document}